\documentstyle[eqsecnum,preprint,aps,prd]{revtex}

\preprint{DAMTP-1999-183}
\date{\today}
\tighten
\begin{document}
\draft
\def\sqr#1#2{{\vcenter{\hrule height.3pt
      \hbox{\vrule width.3pt height#2pt  \kern#1pt
         \vrule width.3pt}  \hrule height.3pt}}}
\def\square{\mathchoice{\sqr67\,}{\sqr67\,}\sqr{3}{3.5}\sqr{3}{3.5}}
\def\today{\ifcase\month\or
  January\or February\or March\or April\or May\or June\or
  July\or August\or September\or October\or November\or December\fi
  \space\number\day, \number\year}


\title{VSL theories and the Doppler peak}

\author{P. P. Avelino${}^{1,2}$\thanks{
Electronic address: pedro\,@\,astro.up.pt} 
C. J. A. P. Martins${}^{3}$\thanks{Also at C.A.U.P.,
Rua das Estrelas s/n, 4150 Porto, Portugal.
Electronic address: C.J.A.P.Martins\,@\,damtp.cam.ac.uk}
G. Rocha${}^{1}$\thanks{
Electronic address: graca\,@\,astro.up.pt} 
}

\address{${}^1$ Centro de Astrof\'{\i}sica, Universidade do Porto\\
Rua das Estrelas s/n, 4150-762 Porto, Portugal}

\address{${}^2$ Dep. de F{\' \i}sica da Faculdade de Ci\^encias da 
Univ. do Porto\\
Rua do Campo Alegre 687, 4169-007 Porto, Portugal}

\address{${}^3$ Department of Applied Mathematics and Theoretical Physics\\
Centre for Mathematical Sciences, University of Cambridge\\
Wilberforce Road, Cambridge CB3 0WA, U.K.}

\maketitle
\begin{abstract}

{We discuss constraints on so-called `varying speed of light theories'
coming from the physics of the early universe.
Specifically, we consider the position of the first acoustic peak
of the CMB angular power spectrum, showing that the recent determination 
of its position by various CMB experiments, in particular BOOMERanG/NA, 
can be used to constrain the
value of the speed of light $c$ after the epoch of last scattering.

Specifically, we find that for a flat universe and a standard density
of baryonic matter a variation in $c$ of up to $4\%$ is consistent
with the current observations. The position of the Doppler peak is
fairly sensitive to changes in $c$, and future observations should 
dramatically improve this bound. On the other hand, we also find that the
maximum baryonic density allowed in VSL theories is
about $\Omega_B h^2\sim 0.06\Omega_0$. We comment on the significance
of these results.}

\end{abstract} 
\pacs{PACS number(s): 98.80.Cq, 95.30.St}
\newpage

\section{Introduction}
\label{secintro} 

Over the past year or so there has been a growing effort to establish
some sort of link between `fundamental' high-energy physics \cite{polc} and the
more standard cosmological scenario \cite{kolb}.
Although this important task is still
at a very early stage (perhaps analogous to the state of cosmology itself some
20 years ago), some general ideas are already emerging. Among these are
the concepts of dimensional reduction and compactification \cite{bank}.
Since high-energy
theories are best formulated in 10 or 11 dimensions and our own low-energy
world is four-dimensional, it is clear that some sort of dimensional reduction
mechanism (whose precise details are still unknown) will be involved.

When one goes through such a procedure, one usually finds that one or more of
the `constants' of Nature are time and/or space-varying quantities.
This normally arises because the fundamental coupling constants are
associated with the radii of additional dimensions, which are usually
variable. Typical examples of this can be found in multi-dimensional
Kaluza-Klein models \cite{chodos}, superstring theories \cite{yswu}, and in
the so-called `brane world' \cite{brane}.
This fact is in part responsible for the recent sharp increase in the
interest in theories with `varying constants', and in particular in
theories where the speed of light $c$ is
time-dependent \cite{M1,M2,BE,AM,B,BM,AM2,BIM}. These are commonly called
`varying speed of light' (VSL) theories, although this is a bit of a
misnomer \cite{AM2}. It should be said that all of these are
essentially toy models, but they are interesting because in certain (but
arguably rather
specific) circumstances they can solve the standard
cosmological problems \cite{G,L1,AS,L2}. We should also point out that on
the observational side there
are some very tentative hints of a time variation (at redshifts of order
unity) of the fine-structure constant \cite{WFCDB}, but these still require
further confirmation. 

As has been noted before \cite{AM2}, there is no unique way to generalise
General Relativity to accommodate variations in $c$ and/or other constants.
At the `toy-model' level, one must therefore choose some particular set
of postulates which will characterise the theory.
Most notably, a number of standard invariance
principles and conservation laws (such as covariance, Lorentz invariance,
mass and particle number conservation) may or may not hold in the generalised
theory. In a previous paper \cite{AM2} we have shown that solutions to the
standard cosmological problems are not a general feature of varying speed of
light (or indeed any `varying constants') theories.
The fact that the theory proposed by Albrecht, Barrow and
Magueijo \cite{AM,B,BM} solves the cosmological problems should be attributed
to it breaking covariance and Lorentz invariance. We have also proposed an
arguably more `natural' generalisation of General Relativity which allows for
variations in $c$ and $G$ and preserves (at least in some cases) all of the
above invariances, and argued that no such generalisation can solve the
standard cosmological problems.

In this paper we discuss some simple constraints on a varying speed of light
in the early universe. In particular, we investigate the constraints imposed
on the theory of Albrecht, Barrow and Magueijo by the recent determination
of the position of the first Doppler peak in the CMB angular power spectrum.

We should point out that there has been some recent work on constraining
variations of $\alpha$ on cosmological timescales \cite{H,KST,BIR}; these
effectively interpret it as a variation in the electric charge $e$ (No
similar work has been done for variations of the speed of light, although
some possible tests have been proposed \cite{tests}). The main
point made by these authors is that a variation of $\alpha$ alters the
ionization history of the universe, and hence changes the pattern of CMB
fluctuations. The dominant effect is a change in the redshift of recombination,
due to a shift in the energy levels (and in particular the binding energy)
of Hydrogen. The Thomson scattering cross section is also changed for all
particles (being proportional to $\alpha^2$). Increasing $\alpha$ increases
the redshift of last scattering (hence the position of the Doppler peak,
$l_p$) and decreases the high-l damping. A smaller effect (which these authors
ignore) is expected to come from a change in the Helium abundance.

It is claimed \cite{H,KST} that the next generation of CMB experiments
can set constraints of the order of
\begin{equation}\label{const1}
\mid \frac{\Delta\alpha}{\alpha}\mid \le10^{-3} \, ,
\end{equation}
or equivalently
\begin{equation}\label{const2}
\mid \frac{\dot \alpha}{\alpha}\mid \le5\times10^{-13}y^{-1}\,
\end{equation}
at a redshift $z\sim10^3$. Note that the claimed detection
of Webb {\em et al.}
\cite{WFCDB} is
\begin{equation}\label{const3}
\frac{\Delta\alpha}{\alpha}\sim(1.5\pm0.3)\times10^{-5} \, ,
\end{equation}
at redshifts $0.5<z<1.6$, although previous work at higher redshifts (up to
$z\sim3$) \cite{varsh} only finds the constraint
\begin{equation}\label{const4}
\mid \frac{\Delta\alpha}{\alpha}\mid \le1.6\times 10^{-4} \, .
\end{equation}
There is also a well-known nucleosynthesis constraint
\begin{equation}\label{const5}
\mid \frac{\dot \alpha}{\alpha}\mid \le2\times10^{-14}y^{-1}\,
\end{equation}
at redshifts $z\sim10^9-10^{10}$ \cite{cpw} although this is a model-dependent
result, since it relies on a particular (and arguably incorrect) model for the
dependence of the neutron to proton mass difference on $\alpha$. More recent
work \cite{BIR} finds a weaker bound by two orders of magnitude.
Finally, for
reference we also add that the best laboratory limit is \cite{prestage}
\begin{equation}\label{const6}
\mid \frac{\dot \alpha}{\alpha}\mid \le3.7\times10^{-14}y^{-1}\, ,
\end{equation}
while the well-known Oklo constraint \cite{oklo} is
\begin{equation}\label{const7}
\mid \frac{\dot \alpha}{\alpha}\mid \le5\times10^{-17}y^{-1}\, .
\end{equation}

The plan of this paper is as follows. In the following section we briefly
review the VSL model of Albrecht, Barrow and Magueijo, being particularly
careful with the choices of fundamental units. Section III contains the main
results of the paper---after a short review of the relevant CMB physics, we
derive constraints on the speed of light at the epoch of recombination, and
briefly discuss nucleosynthesis. Finally, in section IV we summarize our main
results and discuss future work.

\section{The model}
\label{secmodel}

In the VSL theory proposed by Albrecht, Barrow and
Magueijo \cite{AM,B,BM}, one postulates minimal coupling at
the level of Einstein's
equations. Physically, the most obvious consequence of this choice is
that there should be no ${\dot c}$ term in the Friedmann and Raychaudhuri 
equations which can be written in the usual way
given by
\begin{eqnarray}
{\left({\dot a\over a}\right)}^2&=&{8\pi G\over 3}\rho -{Kc^2\over a^2}
\label{friedman1}\\
{\ddot a\over a}&=&-{4\pi G\over 3}{\left(\rho+3{p\over c^2}\right)}\, .
\label{friedman2}
\end{eqnarray}
However,
this choice also implies the breaking of
covariance and Lorentz invariance, as well as of mass and particle number
conservation if the curvature is non-zero. In this case the 
conservation equation is given by
\begin{equation}\label{friedman4}
\dot\rho+3{\dot a\over a}{\left(\rho+{p\over c^2}\right)}=
-\rho{\dot G\over G}+{3Kc^2\over 4\pi G a^2}{\dot c\over c}.
\end{equation}
Note that this also allows for a time-variation of Newton's constant $G$.

As we pointed out above, there is no unique way to generalise General
Relativity in order to include variations of the speed of light or other
fundamental quantities. Even after imposing the above postulate, there is a
lot of freedom remaining. Physically,
the reason for this is that one can only measure dimensionless combinations of
the fundamental parameters. Hence, different choices of units will lead to
theories that, although in the same class, will have different varying
dimensional quantities.

In the theory of Albrecht and Magueijo \cite{AM} the quantity
\begin{equation}\label{defq}
Q=\hbar/c
\end{equation}
is a constant with
\begin{equation}\label{hpropc}
c \propto \hbar \propto \alpha^{-1/2}\, ,
\end{equation}
where $\alpha$ is the 
fine structure constant. They further assume that the mass of elementary 
particles, such as the electron mass, $m_e$, and the electron charge, $e$, are 
constants. It is straightforward to show from dimensional analysis that there 
is a unique set of standard units of mass, length and time (M,L,T) 
which can be constructed using combinations of the fundamental parameters 
$\hbar$, $m_e$, $e'=e/{\sqrt {4\pi \epsilon_0}}$ and $c$ which satisfy 
these criteria. These are
\begin{equation}\label{crit1}
M=m_e\, ,
\end{equation}
\begin{equation}\label{crit2}
L=Q/m_e\, ,
\end{equation}
\begin{equation}\label{crit3}
T=Q^{3/2}/m_e e'\, .
\end{equation}
Hence, 
by making this choice of units we are able to interpret a variation 
in the fine structure constant $\alpha={e'}^2/\hbar c$ 
as being associated to a change in the speed of light, $c$, and 
the Planck constant $\hbar$.

A very important aspect of this theory is the way in which the 
energy of elementary particles scales with $c$. Given that 
the mass of elementary particles is conserved in this theory their energy 
scales as $E \propto c^2$ which means that for the same mass density 
a larger speed of light will imply a hotter universe---with $T \propto c^2$.
Another crucial aspect of this theory is the way in which the Rydberg energy
\begin{equation}\label{defryd1}
E_R = m_e {e'}^4/ 2 \hbar^2\, ,
\end{equation}
which represents 
the dependence of the atomic levels on the fundamental parameters, 
is going to vary in  this cosmology. Given that $m_e$ and $e'$ are 
both constants, we have that $E_R \propto c^{-2}$. 
Hence, a larger speed of light also implies a smaller ionisation temperature. 

Note that, if the Rydberg energy was instead 
\begin{equation}\label{defryd2}
E_R = m_e {e'}^4 \alpha_+^2/ 2 \hbar^2 \alpha^2\, ,
\end{equation}
with $\alpha_+$ being the present value of the fine structure constant
then $E_R$ would scale with $c$ in 
the same way as the energy of the elementary particles. This would 
imply major modifications to quantum mechanics. We will study the 
implications of these scenarios for the spectrum of CMB anisotropies 
in the next section.

\section{The CMB anisotropy}
\label{secCMB}

The description of the CMB anisotropies in terms of the angular power 
spectrum, 
$C_l$, has proved to be an invaluable method and has become a standard 
procedure for treatment of the temperature fluctuations of the CMB radiation.

Contemporary cosmological models with adiabatic fluctuations predict a 
sequence of peaks on the power spectrum which are generated by acoustic 
oscillations of the photon-baryon fluid at recombination.
Photon pressure resists compression of the fluid by gravitational infall and 
sets up acoustic oscillations. The fluctuations as a function of the 
wavenumber $k$ go as 
$\cos(k c_{s} \eta_{ls})$ at last scattering, where $c_{s}$ is the sound 
speed and $\eta_{ls}$ is the conformal time at recombination. This will 
produce a harmonic series of temperature 
fluctuation peaks, 
with the $m$th peak corresponding to
\begin{equation}\label{defkm}
k_{m}=m \pi /c_{s}\eta_{ls}\, .
\end{equation}
in the case of primordial adiabatic fluctuations.
The critical scale is essentially the sound horizon $c_{s} \eta_{ls}$ at 
last scattering \cite{sugi95,hu,2husugi95,huwhite96}.
Of particular interest is the height and position of the main acoustic 
peak---the so called Doppler peak. In the standard cosmological model 
with adiabatic fluctuations produced during inflation 
the height depends on quantities like 
the baryonic content of the universe ($\Omega_{b}$) and the Hubble constant 
($H_0$), whilst its position depends on the total density of the universe 
($\Omega_{0}$, including the contribution of a cosmological constant), and 
is expected to occur on an angular scale $\sim 1^o$ if $\Omega_0 \sim 1$.
The precise form
of the Doppler peak depends on the nature of the dark matter and the
values of $\Omega_0$, $\Omega_b$ and $H_0$.
The scale $l_p$ of the main peak reflects
the size of the horizon at last scattering of the CMB photons and thus depends
almost entirely \cite{hu,kamio} on the total density of the universe 
according to
\begin{equation}\label{deflpeak}
l_p \propto 1/\sqrt{\Omega_0}\, .
\end{equation}
This relation comes from the conversion of a spatial fluctuation on a distant
surface to an anisotropy on the sky. Hence, in an 
open universe a given scale subtends a smaller angle on the sky 
than in the flat universe, due to the fact 
that photons curve in their geodesics.
In a flat $\Lambda$CDM universe the main acoustic peak is 
located at around $l \sim 220$ with a small dependence on $\Omega_{0}$ 
and $h$ \cite{pedro}. In the case of topological defect models the situation 
is not entirely clear yet \cite{ASWA,BRA,CHM}. 

Here we study the shift in the Doppler peak position which may 
be induced by a phase transition after recombination, at a red-shift $z_*$
when the speed of light changed from $c^-$ to $c^+$ (in the following 
we shall assume that $c^- > c^+$). 
From the discussion in the previous section we see that the recombination 
temperature in this theory is given by
\begin{equation}\label{deftemp1}
T_{ls}=T_{ls}^S(c^+/c^-)^2=T_{ls}^S(\alpha_-/\alpha_+)\, 
\end{equation}
if the Rydberg energy has the usual form (see eqn. (\ref{defryd1})). 
The alternative choice  given in eqn. (\ref{defryd2}) would imply that
\begin{equation}\label{deftemp2}
T_{ls}=T_{ls}^S(c^-/c^+)^2=T_{ls}^S(\alpha_+/\alpha_-)\, .
\end{equation}
Here $T_{ls}^S$ and 
$T_{ls}$ represent respectively the recombination temperature in the 
standard and VSL cosmological models. On the other hand, the temperature 
of the photons at a given redshift, $z > z_*$, is given by
\begin{equation}\label{deftemp3}
T(z)=T_\gamma^0(1+z)(c^-/c^+)^2=T_\gamma^0(1+z)(\alpha_+/\alpha_-)\, ,
\end{equation} 
where $T_\gamma^0$ is the temperature of the 
photons today. These two effects combined 
mean that the redshift of recombination 
is smaller than the standard one,
\begin{equation}\label{defredc}
1+z_{ls} = (1+z_{ls}^S) (c^+/c^-)^4 = (\alpha_-/\alpha_+)^2\, ,
\end{equation} 
for the usual form of Rydberg energy, $E_R$ (see  eqn. (\ref{defryd1})). 
If the alternative choice for the Rydberg Energy 
(see  eqn. (\ref{defryd2})) is used then $z^S_{ls}$ and 
$z_{ls}$ would be equal.
 
There are two main effects which can modify the CMB spectrum, shifting the 
main acoustic peak position. Firstly the comoving distance
to the last scattering surface:
\begin{equation}\label{dls}
d=\int_{t_{ls}}^{t_*} c^- \frac{dt}{a(t)}+
\int_{t_*}^{t_0} c^+ \frac{dt}{a(t)},
\end{equation}  
will be altered due to the larger value of the speed of light prior to the 
phase transition. This effect will tend to shift the peak position to smaller 
angular scales but it will be negligible if the redshift of the phase 
transition, $z_*$, is large enough. However, this will be the only effect 
altering the Doppler peak position in the case of the 
`unusual' choice of the Rydberg energy 
(see  eqn. (\ref{defryd2})). If we want the 
shift in the main acoustic peak position due to this effect to be small then 
the following constraint is easily obtained:
\begin{equation}\label{constraint1}
1+z_* \gg (c^-/c^+)^2 = (\alpha_+/\alpha_-)^2.
\end{equation}
The second effect which shifts the position of the main acoustic peak 
if the Rydberg energy has the usual form (see  eqn. (\ref{defryd1})) 
is the change in the sound horizon at last scattering. This is a consequence 
of two competing effects: the larger Hubble radius and the 
smaller value of the 
sound speed which are due to a smaller value of the red-shift of last 
scattering (see eqn. (\ref{defredc})).

For this purpose let us consider, without loss of generality, a flat 
background Universe. In the matter era the conformal time $\eta$ will be 
proportional to $a^{1/2}$. On the other hand the adiabatic sound speed will 
be
\begin{eqnarray}
c_s=1/\sqrt{3} \quad , \quad a < a_{B\gamma} \label{sound11}\\
c_s\propto a^{-1/2} \quad , \quad a_{B\gamma} < a < a_{ls}\, ,
\label{sound2}
\end{eqnarray}
with $a_{B\gamma}$ corresponding to the 
epoch at which $\rho_{\gamma}= \rho_{B}$ (and $B$ standing for baryons). 
Therefore the sound horizon before recombination will evolve as
\begin{eqnarray}
c_s\eta \propto a^{1/2}\quad ,  \quad a < a_{B\gamma} \label{sound13}\\
c_s\eta = {\rm const.}\quad , \quad a > a_{B\gamma}\, ,
\label{sound4}
\end{eqnarray}
Here we should note that although in the standard 
cosmological model $a_{B\gamma} > a_{ls}$ this does not need to be the case 
for our model due to the change of the redshift of last-scattering.
Thus we see that a smaller value of the redshift of recombination will 
imply a larger sound horizon at last scattering. The angle subtended 
by the sound horizon at last scattering, $\theta_{ls}$, will be larger 
and consequently the scale, $l_{p} \propto 1/\theta_{ls}$, 
of the main peak will be smaller. In conclusion, the main effect induced 
by the change in the red-shift of last scattering is an increase in the size 
of the sound horizon at last scattering, shifting the position of the main 
acoustic peak toward lower values of the multipole $l$. It is 
straightforward to include curvature effects in the previous 
considerations. If $z_* \gg 1$ then $\Omega(z_*)$ 
will be close to unity and the above statements will still apply. 
Hence the effect of curvature will be the same as in the standard scenario, 
namely, to shift the position of the peak towards higher values of the 
multipole $l$ (the effect being proportional to $1 / {\sqrt {\Omega_0}}$).

In refs. \cite{H,KST} the effect of a varying fine-structure constant 
$\alpha$ on the CMBR was studied interpreting this variation as being 
due to a change in the electron charge. In this case the Rydberg energy 
(see equation (\ref{defryd1})) is proportional to $\alpha^2$ and consequently 
the recombination temperature in this theory is given by:
\begin{equation}\label{deftemp4}
T_{ls}=T_{ls}^S (\alpha_- / \alpha_+)^2, 
\end{equation}
Given that, in this case, the evolution of the energy of the background 
photons is not affected, the red-shift of recombination is just given by:
\begin{equation}\label{constraint2}
1+z_{ls} = (1+z_{ls}^S)  (\alpha_- / \alpha_+)^2.
\end{equation}
We see that the amplitude of this effect is exactly the same as that 
described by eqn. (\ref{defredc}). However, the shift of the peak 
position due to the change of the distance to the last scattering surface 
is negligible in this case (if $z_{ls} \gg 1$).

We can use the recent determination of the main peak position by
the BOOMERanG \cite{boomer} experiment of \cite{bond}
\begin{equation}\label{boom}
l_{p}=215 \pm 15
\end{equation} 
to obtain constraints to our model. In everything that follows we will use the
usual choice of the Rydberg energy, as discussed above. We will also assume that
the effect of the change in the distance to the last-scattering surface is
small.
For $a_{B\gamma} \ge a_{ls}$, or equivalently $\Omega_{B}h^{2} \le 0.05 (c^+/c^-)^{4}$, we can express the scale $l_{p}$ as follows:
\begin{equation}\label{constraintnew3}
l_{p} \approx \frac{220}{\sqrt{\Omega_0}}
\sqrt{\frac{1+ z_{ls}}{ 1+z_{ls}^{S}}} = 
\frac{220}{\sqrt{\Omega_0}} \left(\frac{c^+}{c^-}\right)^{2} 
\end{equation} 
Using equation \ref{boom} we obtain:
\begin{equation}
\frac{c^+}{c^-}=(1.00 \pm 0.04) \Omega_{0}^{1/4}\, .
\end{equation}
For $a_{B\gamma} \le a_{ls}$, or equivalently $\Omega_{B}h^{2} \ge 0.05 (c^+/c^-)^{4}$ we obtain the following constraint:
\begin{equation} 
\frac{\Omega_{B} h^{2}}{\Omega_0}=0.05 \pm 0.007.
\end{equation}
Thus we see that in our model the maximum allowed value for $\Omega_{B}h^{2}/ \Omega_{0}$ is $\approx 0.06$. This is because the Doppler peak
constrains the epoch when baryons start to dominate over radiation (through the
effect of the sound horizon, as discussed above), and this obviously depends
on how many baryons there are.

We should note that there is also a possible contribution from a third
effect. It has been suggested (see for example \cite{AM}) that a sudden
pase transition will affect the amplitude of any existing density fluctuations,
and that sub-horizon and super-horizon scales may be affected differently.
The precise meaning of these statements is somewhat unclear, as they can easily
be changed depending on the choice of a `preferred gauge' in the theory.
This is a manifestation of the lack of covariance of this formulation
of the model, and is indeed the most serious problem it faces
(and nobody has addressed it so far). In any case it is clear, simply at an
intuitive level, that the density fluctuations might be affected, perhaps
in a scale-dependent way. If this is true, this might introduce a further
shift in the position of the Doppler peak (which has not been addressed here).

Even though we use the toy-model of Albrecht, Barrow and Magueijo, 
our results should be  valid (perhaps with small modifications) for other VSL
prescriptions. Moreover, since we are discussing variations of speed of light
at recent times, we expect these to be at most of the order
of a few percent (as was confirmed {\em a posteriori}).
Any variations in the density fluctuations are expected to be of this
order of magnitude. Since the current error in the determination of the
position of the Doppler peak is itself of order $10\%$,
such effects are not crucial at this stage, although they should of course be
accounted for when more precise data is available.

\section{Discussion and conclusions}
\label{conclusions}

In this paper we investigated the implications of a change in the value 
of the speed of light at some time between last scattering and today in the 
framework of the theory proposed by Albrecht, Barrow and Magueijo. We have 
shown that this may alter the spectrum of the CMBR in particular by shifting 
the position of the main Doppler peak to larger scales. This 
effect may on one hand complicate cosmological 
parameter determinations with CMB but on the other hand could 
bring structure formation models with isocurvature fluctuations 
(including topological defects) in better agreement with the data. 

We find that for a `standard' cosmological scenario, the best currently
available determination of the position of the Doppler peak (that of the
BOOMERanG/NA flight) is consistent with a variation of up to $4\%$ in the
speed of light after the epoch of recombination, relative to the present value.
On the other hand, less standard scenarios are also admissible, provided that
they obey $\Omega_{B}h^{2}\le 0.06 \Omega_{0}$. Future CMB experiments
should dramatically improve these constraints.

In ref. \cite{BIR} stringent constraints 
on variations of the fine structure constant from nucleosynthesis were 
derived (interpreting the variation in $\alpha$ as being due to a variation 
of the electron charge). Here we should note that, even if these constraints 
could be applied to our model there is always 
the possibility that the variation of the speed of light is 
not monotonic making it possible to satisfy nucleosynthesis constraints and 
still affect the CMBR.
Moreover, the problem is further complicated by the fact that, assuming that
one can successfully include a VSL formulation in the context of some
grand-unified theory, one can expect different charges in the theory to
vary at different rates. In any case, in VSL models important
modifications to standard quantum mechanics may be required and only after
these are  properly
introduced will it be possible to make reliable predictions as far as 
nucleosynthesis is concerned.

\acknowledgements

We would like to thank Paulo Carvalho and Paulo Macedo for 
enlightening discussions.
C.M. and G.R. are funded by FCT (Portugal) under
`Programa PRAXIS XXI', grants nos. PRAXIS XXI/BPD/11769/97 and
PRAXIS XXI/BPD/9990/96 respectively.


\end{document}